\documentclass[twocolumn,amsmath,amssymb,superscriptaddress,showpacs,prb]{revtex4}

\usepackage{graphicx}
\usepackage{amsmath,amssymb,amsthm}

\renewcommand{\vec}[1]{\boldsymbol{#1}}
\newcommand{\beq}{\begin{equation}}
\newcommand{\eeq}{\end{equation}}
\newcommand{\beqa}{\begin{eqnarray}}
\newcommand{\eeqa}{\end{eqnarray}}
\newcommand{\e}{\mathrm{e}}
\newcommand{\w}{\omega}

\newcommand{\ket}[1]{\left| #1 \right\rangle}

\newcommand{\ketbra}[2]{\left|#1\right\rangle\hskip-1mm\left\langle #2\right|}

\newcommand{\kv}{ {\bf k} }

\begin{document}

\title{Coherent and incoherent dynamics in excitonic energy transfer: correlated fluctuations and 
off-resonance effects}
\author{Dara P. S. McCutcheon} \email{dara.mccutcheon@ucl.ac.uk}
\affiliation{Department of Physics and Astronomy, University College London, Gower Street, London WC1E 6BT, United Kingdom}
\affiliation{London Centre for Nanotechnology, University College London}
\author{Ahsan Nazir} \email{a.nazir@imperial.ac.uk}
\affiliation{Department of Physics and Astronomy, University College London, Gower Street, London WC1E 6BT, United Kingdom}
\affiliation{Blackett Laboratory, Imperial College London, London SW7 2AZ, United Kingdom}

\date{\today}

\begin{abstract}

We study the nature of the energy transfer process within a pair of coupled two-level systems (donor and acceptor) subject to interactions with the surrounding environment. Going beyond a standard weak-coupling approach, we derive a master equation within the polaron representation that allows for investigation of both weak and strong system-bath couplings, as well as reliable interpolation between these two limits. With this theory, we are then able to explore both coherent and incoherent regimes of energy transfer within the donor-acceptor pair. We elucidate how the degree of correlation in the donor and acceptor fluctuations, the donor-acceptor energy mismatch, and the range of the environment frequency distribution impact upon the energy transfer dynamics. In the resonant case (no energy mismatch) we describe in detail how a crossover from coherent to incoherent transfer dynamics occurs with increasing temperature [A. Nazir, Phys. Rev. Lett. {\bf 103}, 146404 (2009)], and we also explore how fluctuation correlations are able to protect coherence in the energy transfer process. We show that a strict crossover criterion is harder to define when off-resonance, though we find qualitatively similar population dynamics to the resonant case with increasing temperature, while the amplitude of coherent population oscillations also becomes suppressed with growing site energy mismatch.

\end{abstract}


\maketitle

\section{Introduction}

A fascinating series of recent experiments demonstrating signatures of quantum coherence in the energy transfer dynamics of a variety of systems~\cite{lee07,engel07, calhoun09, collini09sc,collini09,collini10,panitchayangkoona10,mercer09,womick09} has sparked renewed interest in modeling excitation energy transfer beyond standard methods.~\cite{gilmore05,nazir09,jang08,jang09,prior10,thorwart09,ishizaki09b,kimura07, roden09, nejad10}
This process, which occurs when energy absorbed at one site (the donor) is transferred to another nearby site (the acceptor) via a virtual photon,~\cite{andrews89} is often considered to be incoherent; the result of weak donor-acceptor interactions, treated perturbatively using Fermi's golden rule.~\cite{foerster59,dexter52} However, though this approach has proved to be immensely successful when applied in many situations,~\cite{scholes03,beljonne09} accounting for {\it quantum coherence} within the energy transfer dynamics requires an analysis beyond straightforward perturbation theory in the donor-acceptor interaction.

An alternative starting point for investigations into coherent energy transfer is to treat the system-environment interaction as a perturbation instead. Such weak-coupling theories, often referred to as being of Redfield or Lindblad type depending upon the approximations made in their derivation,~\cite{b+p} have been successfully applied to elucidate a number of effects that could be at play in multi-site donor-acceptor complexes. Examples include studying the interplay of coherent dynamics and dephasing in promoting efficient energy transfer in quantum aggregates,~\cite{olayacastro08,mohseni08,plenio08,caruso09,rebentrost09,rebentrost09b,chin10} exploring the role of environmental correlations in tuning the energy transfer process,~\cite{fassioli10} and extensions to assess the potential importance of non-Markovian effects.~\cite{renger02,rebentrost09NM}

Nevertheless, in order to properly understand the transition from coherent to incoherent energy transfer which occurs as the system-environment coupling or temperature is increased,~\cite{rackovsky73,leegwater96,gilmore06,nazir09} it is necessary to be able to describe the system dynamics 
beyond either of these limiting cases.~\cite{beljonne09, cheng09,olayacastro10} Building on earlier work,~\cite{soules71,rackovsky73, kenkre74,abram75} a number of methods have been put forward to accomplish this. For example, modifications to both Redfield~\cite{zhang98,yang02, renger03,novoderezhkin04} and F\"orster~\cite{sumi99,scholes00,jang04} theory have extended the range of validity of both approaches. Moreover, it is possible to define a new perturbation term through the small 
polaron transformation,~\cite{wurger98} which under certain conditions allows interpolation between the Redfield and F\"orster limits.~\cite{jang08,nazir09,jang09} 
For particular forms of system-environment interaction, this can also be achieved through the hierarchical equations of motion technique.~\cite{ishizaki09b,tanaka10} Numerically exact calculations, based, for example, on path integral,~\cite{thorwart09,nalbach10} numerical renormalisation group~\cite{tornow08} and density matrix renormalisation group~\cite{prior10} methods, have also been applied to study energy transfer beyond perturbative approaches.

In this work, we investigate the conditions under which coherent or incoherent motion is expected to dominate the energy 
transfer dynamics of a model donor-acceptor pair. Following Ref.~\onlinecite{nazir09}, we employ a Markovian master equation 
derived within the polaron representation for this purpose, 
since it allows for a consistent analysis of the dynamics from weak to strong system-bath coupling 
(or, equivalently, low to high temperatures).~\cite{wurger98,rae02} In addition to presenting a full derivation of the theory, 
we also extend it to explore in detail the important effects of donor-acceptor energy mismatch, deriving 
analytical forms for the dissipative dynamics valid over a large range of parameter space.
Furthermore, we move beyond the scaling limit studied in Ref.~\onlinecite{nazir09} to consider 
an environment frequency distribution of finite extent, characterised by a high-frequency cut-off in the bath spectral density. In the resonant case (no energy mismatch) we define a strict crossover temperature above which the energy transfer dynamics ceases to be coherent.~\cite{nazir09} 

Of particular practical interest is the role played by correlations between the donor and acceptor environmental fluctuations, 
suggested 
as one mechanism by which quantum coherence may survive in the energy transfer process under otherwise adverse conditions~\cite{nazir09,collini09sc,hennebicq09,lee07,yu08,nalbach10, chen10, west10,sarovar09,womick09} (though see, for example, Refs.~\onlinecite{olbrich10} and~\onlinecite{abramavicius10} for alternatives). These correlations are also easily treated within our formalism, through position-dependent couplings between the system and the common environment. As the donor and acceptor are brought closer together, there comes a point at which their 
separation becomes comparable to, or smaller than, the wavelength of relevant modes in the bath. As this happens, fluctuations at each site become ever more correlated, and dephasing effects are suppressed. We shall show, consequently, that as the level of correlation increases, so too does the crossover temperature to the incoherent regime. Hence, strong correlations lead to the survival of coherence at high temperatures. 

Off-resonance, we find that it is less straightforward to define a crossover temperature. 
In contrast to the resonant case, for sufficient energy mismatch between the donor and acceptor, 
increasing the temperature causes the amplitude of the coherent contribution to decrease, 
though not to disappear altogether. In principle, it then becomes possible for a coherent component to exist 
in the dynamics at all but infinite temperatures. Although we are then unable to define a crossover in 
quite the same way, we still find that bath correlations have a qualitatively similar effect to the resonant case, 
protecting coherence in the transfer process. 

The paper is organised as follows. In Section~{\ref{master_equation}} we introduce our model, and derive a master equation describing the donor-acceptor dynamics within the polaron representation. Section~{\ref{resonant_transfer}} considers the resonant case and the coherent-incoherent crossover. 
In Section~{\ref{off_resonance}} we investigate 
off-resonant energy transfer and obtain analytic expressions for the dynamics in a number of limits. Finally, in Section~{\ref{summary}} we summarise our results. 

\section{Polaron transform master equation}
\label{master_equation}

\subsection{The system and polaron transformation}

We consider a donor-acceptor pair ($j=1,2$), each site of which is modeled as a two-level system
with ground state $\ket{G}_j$, excited state $\ket{X}_j$, and energy splitting $\epsilon_j$. 
The pair interact via Coulombic energy transfer with strength $V$, which is responsible for the transfer of excitation  
from one site to the other. We label the state corresponding to a single excitation on site $1$ 
as $\ket{1}\equiv\ket{XG}$, and that on site $2$ as $\ket{2}\equiv\ket{GX}$. 
The environment surrounding the donor-acceptor pair is modelled as a common bath of harmonic oscillators, coupled linearly
to the excited state of each site. The total system-bath Hamiltonian in the single 
excitation subspace (in which energy transfer occurs) is therefore written (where $\hbar=1$) 
%
%
\begin{eqnarray}
H_{\rm SUB}&{}={}&\epsilon_1\ketbra{1}{1}+\epsilon_2\ketbra{2}{2}+V(\ketbra{1}{2}+\ketbra{2}{1})\nonumber\\
&&\:{+}\ketbra{1}{1}B_z^{(1)}+\ketbra{2}{2}B_z^{(2)}+\sum_\kv \w_\kv b_\kv^{\dagger}b_\kv,\nonumber\\
\label{eqn:HSUB}
\end{eqnarray}
where the bath is described by creation (annihilation) operators $b_{\bf k}^{\dagger}$ ($b_{\bf k}$) with corresponding 
angular frequency $\w_{\bf k}$, and wavevector ${\bf k}$. The bath operators are 
given by $B_z^{(j)}=\sum_\kv(g_\kv^{(j)} b_\kv^{\dagger}+g_\kv^{(j)*}b_\kv)$, with coupling 
constants $g_{\bf k}^{(j)}$. As in Ref.~\onlinecite{nazir09}, we shall consider the case in which each site 
is coupled to the bosonic bath with the same magnitude $|g_{\bf k}|$, but make 
the separation between the sites explicit through position-dependent phases in the coupling constants of the form 
$g_{\bf k}^{(j)}=|g_{\bf k}|\e^{i{\bf k}\cdot{\bf r}_j}$, with ${\bf r}_j$ being the position of site $j$. As we shall see, 
this form of coupling gives rise to correlations between the bath influences experienced at each site, 
allowing a range of totally correlated, partially correlated, and completely uncorrelated fluctuations to be explored.~\cite{nazir09, nalbach10} 


A standard weak-coupling approach to the system dynamics would 
now be to derive a master equation for the evolution of the reduced system density operator under the assumption that the system-bath interaction terms, as written in Eq.~(\ref{eqn:HSUB}), can be treated as weak 
perturbations.~\cite{b+p, ishizaki09} In this work, we shall instead derive a master equation describing the donor-acceptor energy transfer dynamics in the (now widely used) polaron representation,~\cite{abram75,nazir09,jang08,jang09,rae02,wurger98,nitzan} whereby we displace the bath oscillators depending on the system state. We may then identify alternative perturbation terms, which can be small over a much larger range of parameter space than those in the original representation. In particular, the polaron framework allows us to reliably explore from weak (single-phonon) to strong (multiphonon) coupling regimes between the system and the bath, provided that the energy transfer interaction $V$ does not become the largest energy scale in the problem (in which case the full polaron displacement is no longer appropriate~\footnote{To ensure this we keep $V<\omega_c$, where $\omega_c$ is a high-frequency cut-off in the bath spectral density (see Eq.~(\ref{spectral_density})). In fact, a rough validity criterion can be given as $(V/\omega_c)^2(1-B^4)\ll1$, see Ref.~\onlinecite{mccutcheon10}, which for small enough $V/\omega_c$ is satisfied regardless of the size of the system-bath coupling strength or temperature.}), and that there is no infra-red divergence in its bath-renormalised value $V_R$ (see Eq.~(\ref{B_integral}) below).~\cite{silbey84} In contrast, with a weak system-bath coupling treatment we would only be able to probe single-phonon bath-induced processes, and hence not be able to properly explore the crossover from coherent to incoherent dynamics in which we are primarily interested.

To proceed, we thus apply a unitary transformation which displaces the bath oscillators according to the location of the 
excitation. Defining 
$H_P=\e^{S}H_{\rm SUB}\e^{-S}$, where 
\beq
S=\ketbra{1}{1} P(g_\kv^{(1)}/\w_\kv)+\ketbra{2}{2} P(g_\kv^{(2)}/\w_\kv),
\eeq
with bath operators $P(\alpha_\kv)=\sum_\kv(\alpha_\kv b_\kv^{\dagger}-\alpha_\kv^* b_\kv)$, 
results in the polaron transformed spin-boson Hamiltonian~\cite{wurger98} $H_P=H_0+H_I$, 
with
\beq
H_0=\frac{\epsilon}{2}\sigma_z+V_R\sigma_x+\sum_\kv\w_\kv b_\kv^{\dagger}b_\kv,
\label{eqn:HP}
\eeq
and 
\beq
H_I=V(B_x\sigma_x+B_y\sigma_y).
\label{H_I}
\eeq
Here, the bias $\epsilon=\epsilon_1-\epsilon_2$, gives the energy difference between the donor and acceptor, while the Pauli operators are defined in a basis in which $\sigma_z=|1\rangle\langle1|-|2\rangle\langle2|=|XG\rangle\langle XG|-|GX\rangle\langle GX|$. The bath operators appearing in Eq.~({\ref{eqn:HP}}) are constructed as $B_x=(1/2)(B_++B_--2B)$ and $B_y=(i/2)(B_+-B_-)$, where 
\beq
B_{\pm}=\prod_\kv D\Bigg(\pm\frac{(g_\kv^{(1)}-g_\kv^{(2)})}{\w_\kv}\Bigg),
\eeq
with displacement operators $D(\pm \alpha_\kv)=\exp[\pm(\alpha_\kv b_\kv^{\dagger}-\alpha_\kv^* b_\kv)]$. Note that the interaction terms in Eq.~(\ref{H_I}) therefore depend upon the difference in donor and acceptor system-bath couplings $g_{\bf k}^{(1)}$ and $g_{\bf k}^{(2)}$, respectively.
Importantly, the term
driving coherent energy transfer in Eq.~(\ref{eqn:HP}) will not be treated perturbatively, though it does now have a bath-renormalised strength, $V_R=BV$,
where 
\beq
B=\exp\bigg[-\sum_\kv \frac{|g_\kv|^2}{\w_\kv^2}(1-\cos(\kv \cdot{\bf d}))\coth(\beta \w_\kv/2)\bigg]
\label{B_summation}
\eeq
is the expectation value of the bath operators with respect to 
the free Hamiltonian: $B=\langle B_{\pm} \rangle_{H_0}$. The donor-acceptor separation is given by ${\bf d}={\bf r}_1-{\bf r}_2$.

In order to calculate the renormalisation factor, we take the continuum limit to convert the summation in Eq.~({\ref{B_summation}}) into an integral.
Defining the bath spectral density $J(\w)=\sum_\kv |g_\kv|^2\delta(\w-\w_\kv)$, which contains 
information regarding both the density of oscillators in the bath with a given frequency, and also how strongly those
oscillators interact with the donor-acceptor pair, and assuming a linear, isotropic dispersion relation, we find
\beq
B=\exp\bigg[-\int_0^{\infty}\frac{J(\w)}{\w^2}(1-F_D(\w,d))\coth(\beta \w/2)\bigg].
\label{B_integral}
\eeq
Here, $\beta=1/k_BT$ is the inverse temperature, while the function $F_D(\w,d)$ captures the degree of spatial correlation in the bath fluctuations seen at each site, and 
is dependent upon the dimensionality of the system-bath interaction ($D=1,2,3$).~\cite{mccutcheon09,fassioli10,nalbach10} 
In one dimension $F_1(\w,d)=\cos(\w d/c)$, with $c$ the bosonic excitation speed, in two dimensions $F_2(\w,d)=J_0(\w d/c)$, where $J_0(x)$ is a Bessel function of the first kind, and in three dimensions $F_3(\w,d)=\mathrm{sinc}(\w d/c)$. In all cases $F_D(\w,d)\rightarrow 1$ as $d\rightarrow 0$, i.e. when the donor and acceptor are at the same position, bath fluctuations are perfectly correlated, and the energy transfer strength is not renormalised ($V_R\rightarrow V$). In fact, in this limit dissipative process 
are entirely suppressed (provided $|g_{\bf k}^{(1)}|=|g_{\bf k}^{(2)}|$) and energy transfer remains coherent for all times and in all parameter regimes in our model (the single-excitation subspace is then decoherence-free.~\cite{lidar98, zanardi97}) 
In two and three dimensions, as $d\rightarrow \infty$, $F_D(\w,d)\rightarrow 0$, and the renormalisation takes on the value that would be obtained by considering separate, completely uncorrelated baths surrounding the donor and acceptor. In the following, we shall characterise the degree of correlation in terms of the dimensionless parameter $\mu=c/\omega_0d$, where $\omega_0$ is a typical bath frequency scale (see Eq.~(\ref{spectral_density}) below). We therefore have $\mu=0$ in the absence of correlations, $\mu<1$ for weak correlations, and $\mu>1$ for strong correlations. 

\subsection{Markovian master equation}

Having identified a new perturbation term by transforming our Hamiltonian to the polaron representation, we can now construct a master equation describing the evolution of the donor-acceptor pair reduced density operator $\rho$ up to second order in $H_I$. We employ a standard Born-Markov approach, which yields a polaron frame, interaction picture master equation of the form~\cite{b+p}
\beq
\frac{\partial\tilde{\rho}(t)}{\partial t}=-\int_0^{\infty}d\tau\,\mathrm{tr}_B\big[[\tilde{H}_I(t),[\tilde{H}_I(t-\tau),\tilde{\rho}(t) \otimes \rho_B]\big],
\label{master_equation_1}
\eeq
where tildes indicate operators in the interaction picture, $\tilde{O}(t)=\e^{i H_0 t}O\e^{-i H_0 t}$, and $\mathrm{tr}_B$ denotes a trace over the bath degrees of freedom. In deriving Eq.~({\ref{master_equation_1}}) we have assumed: (i) factorising initial conditions for the joint system-bath density operator within the polaron frame, $\chi(0)=\rho(0)\otimes \rho_B$, with $\rho_B=e^{-\beta H_B}/{\rm tr}_B(e^{-\beta H_B})$ 
being a thermal equilibrium state of the bath; (ii) that by construction the interaction is weak in the polaron frame so that we may factorise the joint density operator as $\tilde{\chi}(t)=\tilde{\rho}(t)\rho_B$ at all times; (iii) that the timescale on which the donor-acceptor system evolves appreciably in both the Schr\"{o}dinger and interaction pictures is large compared to the bath memory time $\tau_B$. Since, for the spectral density we shall consider below, $\tau_B\sim1/\omega_c$, where $\omega_c$ is a high-frequency cutoff (see Eq.~(\ref{spectral_density})), this is not too restrictive, as we must keep $V<\omega_c$ anyway in order for the polaron theory to work well. We note that interesting non-Markovian and non-equilibrium bath effects have been explored in the polaron formalism in Refs.~\onlinecite{jang08} and~\onlinecite{jang09}.

Inserting Eq.~({\ref{H_I}}) into Eq.~({\ref{master_equation_1}}), and moving back into the Schr\"{o}dinger picture, we arrive at our Markovian master equation describing the energy transfer dynamics within the single-excitation subspace, and written in the polaron frame as
\beq
\begin{split}
\frac{\partial\rho(t)}{\partial t}=&-i[(\epsilon/2)\sigma_z+V_R\sigma_x,\rho(t)]\\
&-V^2\int_0^\infty \mathrm{d}\tau\Bigl([\sigma_x,\tilde{\sigma}_x(-\tau)\rho(t)]\Lambda_{xx}(\tau)\\
&\hspace{1.7cm}+[\sigma_y,\tilde{\sigma}_y(-\tau)\rho(t)]\Lambda_{yy}(\tau)+\mathrm{H.c.}\Bigr),
\label{master_equation_2}
\end{split}
\eeq
where H.c. denotes Hermitian conjugation. 
The effect 
of the bath is now contained within the correlation functions 
$\Lambda_{ll}(\tau)=\langle \tilde{B}_l(\tau)\tilde{B}_l(0)\rangle_{H_0}$, which are given explicitly by
\begin{align}
\Lambda_{xx}(\tau)&=(B^2/2)(\e^{\phi(\tau)}+\e^{-\phi(\tau)}-2)\label{Lambda_xx},\\
\Lambda_{yy}(\tau)&=(B^2/2)(\e^{\phi(\tau)}-\e^{-\phi(\tau)})\label{Lambda_yy},
\end{align}
where
\beq
\begin{split}
\phi(\tau)=2\int_0^{\infty}d\w&\bigg[\frac{J(\w)}{\w^2}(1-F_D(\w,d))\\
\times&\left(\cos \w\tau\coth(\beta\w/2)-i \sin \w\tau\right)\bigg].
\label{phi}
\end{split}
\eeq
Notice that the phonon propagator, $\phi(\tau)$, is correlation-dependent due to the factor $(1-F_D(\w,d))$, and so clearly the dissipative effect of the bath will be dependent upon the degree of correlation too. For example, as $d\rightarrow0$, $F_D(\w,d)\rightarrow 1$, and the dissipative contribution 
to Eq.~({\ref{master_equation_2}}) vanishes, as anticipated earlier. 

\subsection{Evolution of the Bloch vector}
\label{evolution_of_the_bloch_vector}

We solve our master equation in terms of the Bloch vector, defined as 
$\vec{\alpha}=(\alpha_x,\alpha_y,\alpha_z)^T=(\langle\sigma_x \rangle, \langle\sigma_y \rangle, \langle\sigma_z \rangle)^T$. As we are working exclusively in the single-excitation subspace, $\alpha_x$ and $\alpha_y$ describe the coherences between the states $|1\rangle\equiv|XG\rangle$ and $|2\rangle\equiv|GX\rangle$, while $\alpha_z$ captures the donor-acceptor population transfer dynamics generated by the coupling $V$. 

Though Eq.~(\ref{master_equation_2}) is written
in the Sch\"{o}dinger picture, it is still in the polaron frame, and so we must determine how expectation values 
in the polaron frame are related to those in the original, or ``lab" frame. We can see this by writing 
${\alpha}_i=\mathrm{tr}_{S+B}(\sigma_i{\chi}_L(t))=\mathrm{tr}_{S+B}(\sigma_i \e^{-S}{\chi}(t)\e^S)
=\mathrm{tr}_{S+B}(\e^S \sigma_i e^{-S}{\rho}(t)\rho_{B})$, 
where $\chi_L(t)=e^{-S}\chi(t)e^S$ is the lab frame total density operator, and we have made use of the Born approximation in the polaron frame to write $\chi(t)=\rho(t)\rho_{B}$. Since $\e^S \sigma_x \e^{-S}=\ketbra{2}{1} B_- +\ketbra{1}{2}B_+$, 
$\e^S \sigma_y \e^{-S}=i(\ketbra{2}{1} B_- -\ketbra{1}{2}B_+)$, and $\e^S \sigma_z \e^{-S}=\sigma_z$, this implies that 
the lab Bloch vector elements are ${\alpha}_i=B{\alpha}_{iP}$, for $i=x,y$, and ${\alpha}_z={\alpha}_{zP}$, where $\alpha_{iP}$ is an expectation value in the polaron frame: 
${\alpha}_{iP}=\mathrm{Tr}_S(\sigma_i{\rho}(t))$. Alternatively, we can define a matrix $L$ which maps the polaron frame Bloch vector ($\vec{\alpha}_P$) to its lab frame counterpart ($\vec{\alpha}$): $\vec{\alpha}=L \cdot \vec{\alpha}_P$, where $L=\mathrm{diag}(B,B,1)$.

Working in terms of the Bloch vector, we arrive at an equation of motion of the form
\beq\label{labbloch}
\dot{\vec{\alpha}}(t)=M\cdot\vec{\alpha}(t)+\vec{b}.
\eeq
In the following, we shall often be interested in determining whether the energy transfer dynamics is predominantly coherent or incoherent. It is then helpful to write
Eq.~({\ref{labbloch}}) as 
\beq
\dot{\vec{\alpha}}'(t)=M\cdot\vec{\alpha}'(t),
\eeq
with $\vec{\alpha}'(t)=\vec{\alpha}(t)-\vec{\alpha}(\infty)$, where 
$\vec{\alpha}(\infty)=-M^{-1}\cdot \vec{b}$ is the steady state. 
This makes clear that the nature of the energy transfer process 
lies solely in the matrix $M$, 
while the inhomogeneous term $\vec{b}$ is needed only in determining the steady state. 

Equipped with the eigensystem of $M$, we may determine the corresponding time evolution as follows: an eigenvector of $M$, say $\vec{m}_i$, has equation 
of motion $\dot{\vec{m}}_i=q_i\vec{m}_i$, where $q_i$ is the corresponding eigenvalue. Its subsequent evolution 
then has the simple exponential form $\vec{m}_i(t)=\vec{m}_i\e^{q_i t}$. More generally, we can say that any initial state 
$\vec{\alpha}'(0)$ will have subsequent evolution 
\beq
\vec{\alpha}'(t)=\sum_{i=1}^{3} a_i \vec{m}_i \e^{q_i t},
\label{general_evolution}
\eeq
where the coefficients $a_i$ are determined by the initial conditions (i.e. the solutions of 
$\vec{\alpha}'(0)=\sum_i a_i \vec{m}_i$). 
The solution to the full inhomogeneous 
equation is then found simply by addition of the steady state: 
$\vec{\alpha}(t)=\vec{\alpha}'(t)+\vec{\alpha}(\infty)$.

\section{Resonant energy transfer}
\label{resonant_transfer}

We start by considering the important special case of resonant energy transfer, in which the interplay of coherent and incoherent effects is particularly pronounced. As we shall see, in this situation it is relatively straightforward to derive a strict criterion governing whether or not we expect the energy transfer dynamics to be able to display signatures of coherence.~\cite{nazir09,wurger98} Hence, resonant conditions provide a natural situation in which to begin to understand, for example, the role of bath spatial correlations~\cite{nazir09, nalbach10, chen10, fassioli10, yu08, hennebicq09,west10,sarovar09,womick09} or the range of the bath frequency distribution in determining the nature of the energy transfer process.

Setting the donor-acceptor energy mismatch to zero,
$\epsilon=0$, we find from Eq.~(\ref{master_equation_2}) dynamics generated by an expression of the form 
$\dot{\vec{\alpha}}=M_R\cdot\vec{\alpha}+\vec{b}_R$, with 
\beq
M_R=\left(
\begin{array}{ccc}
-(\Gamma_z-\Gamma_y) & 0 & 0 \\
0 & -\Gamma_y & -2 B V_R \\
0 & B^{-1}(2 V_R+\lambda_3) & -\Gamma_z
\end{array}
\right),
\label{resonant_M}
\eeq
and $\vec{b}_R=(-B \kappa_x,0,0)^T$, where
\begin{align}
\Gamma_y&=2V^2\gamma_{xx}(0),\\
\Gamma_z&=V^2(\gamma_{yy}(2V_R)+\gamma_{yy}(-2V_R))+2V^2\gamma_{xx}(0),\\
\lambda_3&=2V^2(S_{yy}(2V_R)-S_{yy}(-2V_R)),\\
\kappa_x&=V^2(\gamma_{yy}(2V_R)-\gamma_{yy}(-2V_R)).
\end{align}
The rates and energy shifts are related to the response functions 
\beq
K_{ii}(\w)=\int_0^{\infty}\mathrm{d}\tau\e^{i \w \tau}\Lambda_{ii}(\tau)\mathrm=\frac{1}{2}\gamma_{ii}(\w)+i S_{ii}(\w),
\eeq
such that
\beq
\gamma_{ii}(\w)=2\mathrm{Re}[K_{ii}(\w)]=\int_{-\infty}^{+\infty}\mathrm{d}\tau\e^{i \w \tau}\Lambda_{ii}(\tau),
\eeq
and $S_{ii}(\w)=\mathrm{Im}[K_{ii}(\w)]$.

The resonant steady-state is straightforwardly found to be
\beq
\alpha_x(\infty)=- B \tanh(\beta V_R),
\eeq
while $\alpha_y(\infty)=\alpha_z(\infty)=0$. Notice that while this is similar in form to the steady-state that would be obtained from a weak system-bath coupling treatment,~\cite{b+p, weissbook} $\alpha_x(\infty)$ is determined here by $V_R$, rather than the original coupling $V$, and there is also an extra factor of $B$ suppressing its magnitude. 

The eigenvalues of $M_R$ are given by 
$q_1=\Gamma_y-\Gamma_z$ and $q_2=q_3^*=-(1/2)(\Gamma_y+\Gamma_z+i \xi_R)$. 
Thus, referring to Eq.~(\ref{general_evolution}), we see that
\beq
\xi_R=\sqrt{8V_R(2V_R+\lambda_3)-(\Gamma_z-\Gamma_y)^2}
\label{oscillation_frequency}
\eeq
determines whether or not any coherence exists within the energy transfer dynamics.
Considering the initial state $\vec{\alpha}(0)=(0,0,1)^T$, corresponding to excitation of the donor, $\rho(0)=|1\rangle\langle1|=|XG\rangle\langle XG|$, we find analytical forms for the evolution of the Bloch vector components:
\begin{eqnarray}
\alpha_x(t)&=&-B\tanh(\beta V_R)(1-\e^{-(\Gamma_y-\Gamma_z)t}),\\
\alpha_y(t)&=&-\frac{2 B V_R}{\xi_R}\e^{-(\Gamma_y+\Gamma_z)t/2}\sin\Big(\frac{\xi_R t}{2}\Big),\label{y_general_evolution}\\
\alpha_z(t)&=&\e^{-(\Gamma_y+\Gamma_z)t/2}\Bigl[\cos\Bigl(\frac{\xi_R t}{2}\Bigr) 
+\frac{\Gamma_y-\Gamma_z}{\xi_R}\sin\Bigl(\frac{\xi_R t}{2}\Bigr)\Bigr].\label{z_general_evolution}\nonumber\\
\end{eqnarray}
Inspection of Eqs.~(\ref{oscillation_frequency}) and ({\ref{z_general_evolution}}) 
allows us to identify a crossover from coherent to incoherent motion in the energy transfer dynamics as the point 
at which oscillations in the population difference vanish:~\cite{nazir09}
\beq
(\Gamma_z-\Gamma_y)^2=8V_R(2V_R+\lambda_3).
\label{critical_condition}
\eeq
For $(\Gamma_z-\Gamma_y)^2<8V_R(2V_R+\lambda_3)$, $\xi_R$ is real and both the 
population difference and coherence $\alpha_y$ describe damped oscillations, while for $(\Gamma_z-\Gamma_y)^2\geq8V_R(2V_R+\lambda_3)$, $\xi_R$ is either zero or imaginary, with the resulting dynamics then being entirely incoherent.

\begin{figure}[!t]
\begin{center}
\includegraphics[width=0.45\textwidth]{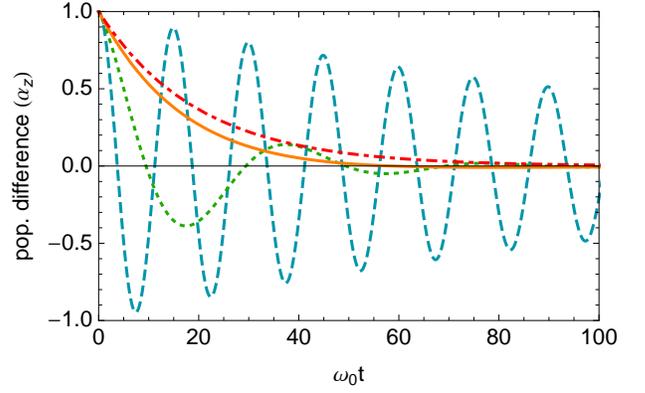}
\caption{Population difference as a function of scaled time $\w_0 t$ for temperatures 
of $k_B T/\w_0=1$ (blue dashed curve), $k_B T/\w_0=5$ (green dotted curve),  $k_B T/\w_0=12$ (orange solid curve) and $k_B T/\w_0=20$ (red dot-dashed curve). Parameters: $\alpha=0.05$, $V/\w_0=0.5$, 
$\w_c/\w_0=4$, $\epsilon=0$ and $\mu=c/\omega_0d=0.5$.} 
\label{e-zero_dynamics}
\end{center}
\end{figure}

\begin{figure}[!t]
\begin{center}
\includegraphics[width=0.45\textwidth]{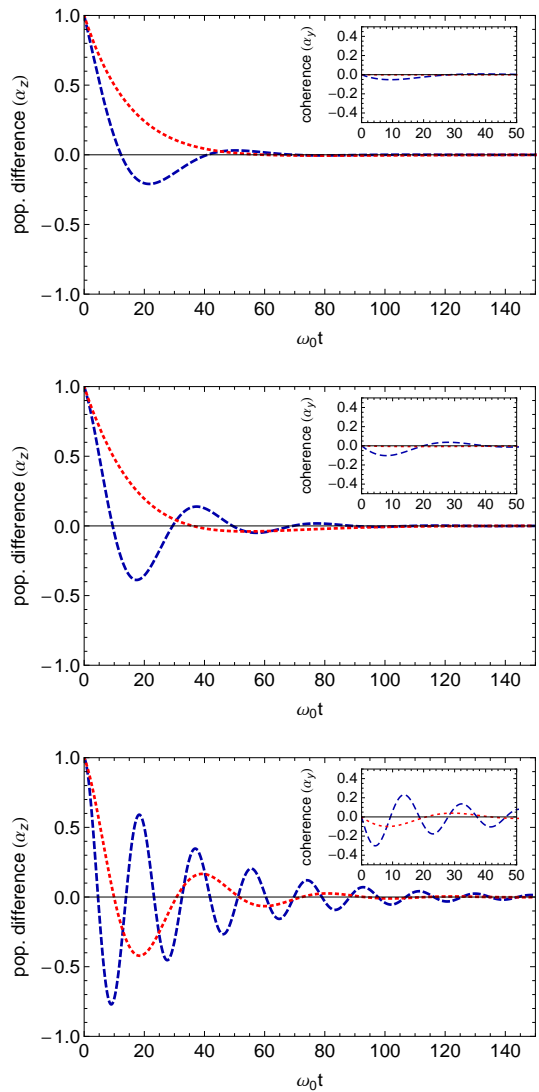}
\caption{Population difference as a function of scaled time $\w_0 t$ for temperatures 
of $k_B T/\w_0=5$ (blue dashed curves) and $k_B T/\w_0=10$ (red dotted curves), and for separations 
corresponding to no correlation, $\mu=c/\w_0d=0$ (top), weak correlations, $\mu=0.5$ (middle), 
and strong correlations $\mu=2$ (bottom). The insets show the evolution of the 
corresponding coherence $\alpha_y$. Parameters: $\alpha=0.05$, $V/\w_0=0.5$, 
and $\w_c/\w_0=4$.} 
\label{resonant_dynamics}
\end{center}
\end{figure}

To further analyse the behaviour of $\alpha_z(t)$, and the 
conditions for which the boundary defined by Eq.~({\ref{critical_condition}}) is crossed, we now take a 
specific form for the system-bath spectral density. For a large enough bath we may approximate 
$J(\w)=\sum_\kv |g_\kv |^2\delta(\w_\kv-\w)$ as a smooth function of $\w$. In this work we consider a spectral 
density of the form
\beq
J(\w)=\alpha \frac{\w^3}{\w_0^2} \e^{-\w/\w_c},
\label{spectral_density}
\eeq
where $\alpha$ is a dimensionless quantity capturing the strength of the system-bath interaction, and
$\w_0$ is a typical frequency of bosons in the bath, which sets an overall energy scale. The cubic frequency dependence in Eq.~({\ref{spectral_density}}) is typical, for example, 
in describing dephasing due to coupling to acoustic phonons,~\cite{wurger98,ramsay10} but can also be used to elucidate the behaviour in which we are interested in general.~\cite{nalbach10} The cut-off frequency $\w_c$ 
is needed to ensure that vacuum contributions remain finite, and is 
related to parameters specific to the particular physical system one wishes to model. The inverse cut-off frequency also sets a typical relaxation timescale for the bath.~\cite{b+p}

To illustrate the dynamics and crossover behaviour in the resonant case, in Fig.~\ref{e-zero_dynamics} we plot the population difference ($\alpha_z$) as a function of the scaled time $ \w_0 t$ for a range of temperatures, showing the transition from coherent 
to incoherent transfer as the temperature is increased. In this plot, and all the following, we consider three-dimensional coupling, $F_3(\omega,d)={\rm sinc}(\omega d/c)$. 
The role of bath spatial correlations in protecting coherence can be seen in Fig.~{\ref{resonant_dynamics}}, where we again plot the evolution of the population difference (the insets show the corresponding coherence $\alpha_y$), this time for representative intermediate and high-temperature cases. The different plots in Fig.~\ref{resonant_dynamics} correspond to zero correlations, characterised by $\mu=c/\w_0d=0$ ($d\rightarrow\infty$, top), weak correlations, $\mu=0.5$ (middle), and strong correlations $\mu=2$ (botttom).~\footnote{Although we would usually expect $V$ to change with varying separation $d$, we keep it fixed for all plots presented here in order to isolate the role played by the environmental correlations.} Progressing from the uppermost plot to the lowest, we clearly see that an increase in correlation strength 
prolongs the timescale over which oscillations in both the 
population difference and coherence persist. Moreover, by looking at the curves corresponding to the 
higher temperature (red, dotted), we can see that as the degree of correlation is 
increased from zero, the dynamics moves from a regime showing purely incoherent relaxation, 
to a regime which displays coherent oscillations {\it at the same temperature}. 
The increase in correlations is thus able to extend the region of parameter space which permits coherence,~\cite{nazir09} as we shall now explore in greater detail. 

\subsection{Coherent to incoherent transition}

We now return our attention to the crossover from 
coherent to incoherent transfer, defined by Eq.~({\ref{critical_condition}}). Intuitively, we might expect the dynamics in the low-temperature (or weak-coupling) regime to be 
coherent; for example, 
in Fig.~{\ref{e-zero_dynamics}} 
incoherent relaxation only occurs in the high-temperature limit. 
If we therefore assume that the crossover itself occurs in the high-temperature regime, it is possible to derive 
an analytic expression governing the crossover temperature by approximating the rates $\Gamma_y$ and $\Gamma_z$. Details of this approximation, and its range of validity, can be found in Appendix~{\ref{high_temperature_rates}}. Generally, for high enough temperatures and/or strong enough system-bath coupling (such that $\beta V_R\ll1$) we can approximate $\gamma_{xx}(\eta)\approx\gamma_{yy}(\eta)\approx\gamma_{yy}(0)$ in $\Gamma_y$ and $\Gamma_z$, where
\beq
\gamma_{yy}(0)\approx\frac{\beta B^2\e^{\phi_0 C_0(x,y)}}{2 \sqrt{\pi C_2(x,y) \phi_0}},
\label{strong_rate}
\eeq
with $\phi_0=2 \pi^2 \alpha/\w_0^2\beta^2$, $x=\pi d/c \beta$ and $y=\w_c\beta$. The functions $C_0(x,y)$ 
and $C_2(x,y)$ are given by Eqs.~({\ref{C_0}}) and ({\ref{C_2}}), and the renormalisation factor $B$ by the product of 
Eqs.~({\ref{B_vacuum}}) and ({\ref{B_thermal}}). If we further assume that the 
energy shift $\lambda_3$ vanishes in the high-temperature limit, Eq.~({\ref{critical_condition}}) reduces to
\beq
(\Gamma_z-\Gamma_y)=4 V_R,
\label{critical_condition_2}
\eeq
and we arrive at the expression
\beq
\bigg(\frac{k_B T}{\w_0}\bigg)^2=\frac{V}{\w_0}\frac{B \e^{\phi_0 C_0(x,y)}}{4 \sqrt{2 \pi^3 \alpha C_2(x,y)}},
\label{approx_critical_condition}
\eeq
with solution, $T_c$, giving the crossover temperature separating the coherent and incoherent regimes. 

\begin{figure}[!t]
\begin{center}
\includegraphics[width=0.45\textwidth]{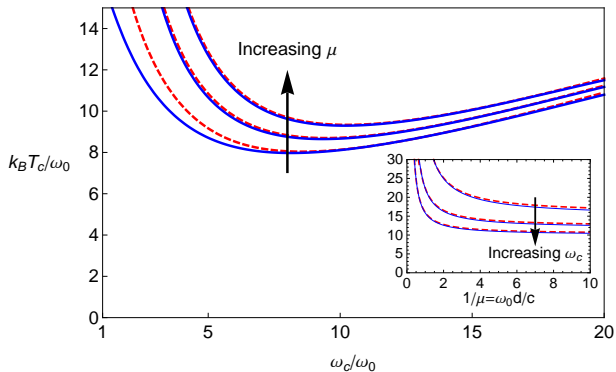}
\caption{Crossover temperature separating the coherent and incoherent regimes against cut-off frequency, for 
levels of correlation given by $\mu= c/\w_0d=0$, $\mu=0.5$ and $\mu=1$, increasing as shown. The solid blue curves have been calculated from Eq.~(\ref{critical_condition}) (using the full rates), while the dashed red curves are solutions to the high-temperature approximation, Eq.~({\ref{approx_critical_condition}}). The inset shows the dependence 
on the level of correlation ($1/\mu=\omega_0d/c$) for different cutoffs, $\w_c/\w_0=2$, $\w_c/\w_0=3$, and $\w_c/\w_0=4$, again increasing as shown. Parameters: $\alpha=0.05$ and $V/\w_0=0.5$.} 
\label{scaling_TC2}
\end{center}
\end{figure}

The dependence of $T_c$ on the various parameters involved in the problem is not straightforward, owing to the temperature dependence in the renormalisation factor $B$, in the functions $C_0$ and $C_2$, and in $\phi_0$. 
In fact, there are three distinct and important temperature scales which determine when coherent or incoherent processes dominate: $T_0=\omega_0/(\sqrt{2\alpha}\pi k_B)$, which depends upon the system-bath coupling strength; $T_x=c/d\pi k_B$, which arises due to the fluctuation correlations and becomes unimportant in the uncorrelated case ($T_x\rightarrow0$ as $d\rightarrow\infty$); and $T_y=\omega_c/k_B$, dependent upon the cut-off frequency, and irrelevant in the scaling limit ($y\rightarrow\infty$). Hence, changes in any of $\alpha$, $d$, or $\omega_c$ can have an effect on the crossover temperature.
For example, the main part of Fig.~{\ref{scaling_TC2}} shows the solution to Eq.~({\ref{approx_critical_condition}}), i.e. the crosover temperature $T_c$, 
as a function of the dimensionless cut-off frequency $\w_c/\w_0$. A calculation using Eq.~(\ref{critical_condition}) with the full rates, and including $\lambda_3$, is also shown for comparison. The three pairs of curves correspond to increasing levels of correlation, ordered as indicated. We see that, except for small $\omega_c/\omega_0$ in the case $\mu=0$, 
where $\lambda_3$ becomes important, solutions to Eq.~(\ref{approx_critical_condition}) give an excellent approximation to the crossover temperature calculated using the full rates. This confirms that the coherent-incoherent crossover does indeed occur in the high-temperature (multiphonon) regime, and consequently could not be captured by a weak system-bath coupling treatment.

As the cut-off frequency is increased from its minimum value, the crossover temperature begins to decrease. This behaviour can be understood qualitatively by examining Eq.~({\ref{critical_condition_2}}), and considering the competition this condition captures between the rate $\Gamma_z-\Gamma_y$ and the coherent interaction $V_R$ in defining the nature of the dynamics. Larger values of the cut-off frequency correspond to 
smaller values of the renormalised interaction strength $V_R$ (see e.g. Eq.~(\ref{B_vacuum})), while 
the rates $\Gamma_y$ and $\Gamma_z$ vary less strongly with $\w_c$ in this regime. Thus, increasing $\w_c$ from its minimum value decreases $V_R$, and therefore reduces the range of temperatures for which $4V_R>\Gamma_z-\Gamma_y$ and coherent transfer can take place. Thus, the crossover temperature falls. Physically, this can be understood by noting that as the cut-off frequency is increased, so too is the effective frequency range and peak magnitude of the system-bath interaction, characterised by the spectral density [Eq.~(\ref{spectral_density})]. Hence, increasing from small $\omega_c/\omega_0$, the environment begins to exert an enhanced influence on the system behaviour, and so coherent dynamics no longer survives to such high temperatures. As $\w_c$ continues to increase, however, we see the crossover temperature then begins to rise. The renormalisation factor $B$ tends to zero with increasing $\w_c$ and here becomes the dominating quantity, thus causing the rate $\Gamma_z-\Gamma_y\sim \mathcal{O}(B^2)$ to vanish faster than the renormalised donor-acceptor coupling $V_R=BV$. 

The interplay between the size of $\w_c$ and the level of spatial correlation is best understood by considering 
the inset of Fig.~{\ref{scaling_TC2}}. For all curves shown the crossover temperature increases as the distance $d$ is reduced, since the level of correlation $\mu$ increases correspondingly. As we have seen previously in Fig.~\ref{resonant_dynamics}, stronger correlations allow coherent dynamics to be observed at higher temperatures; since environmental effects are suppressed, so the crossover temperature $T_c$ must rise. This behaviour can 
be attributed to an increase in the renormalised interaction strength, $V_R$, in relation to the rate $\Gamma_z-\Gamma_y$, this time with variations in the correlation level $\mu$. Interestingly, as the cut-off frequency is increased up to $\w_c/\w_0=4$ (lowest curve), we see that not only does the crossover temperature decrease, but also that the degree of correlation necessary to show a marked rise in $T_c$ increases. As can be seen by comparing the separation between the different curves in the main part of the figure, increasing the cut-off frequency tends to 
suppress the extent to which correlations are able to protect coherence in the system. This tallies with the dynamics shown in Fig.~{\ref{resonant_dynamics}}, for which $\w_c/\w_0=4$, and correlations as high as $\mu=2$ were needed before a significant change in behaviour was seen. Finally, since the renormalisation factor $B$ tends to a constant non-zero value as the correlations vanish at large $d$ (as opposed to $B\rightarrow0$ as $\w_c\rightarrow \infty$), the dependence of the crossover 
temperature on $\mu$ is monotonic, in contrast to its dependence on $\w_c$.

\section{Off resonance}
\label{off_resonance}

It is often the case in practice that the donor and acceptor will have 
different excited state energies, $\epsilon_1-\epsilon_2=\epsilon\neq0$, and so we now turn our attention to energy transfer dynamics under off-resonant conditions. 
Regarding the coherent to incoherent 
transition, in the resonant case we were able to identify this point with a pair of conjugate eigenvalues converging on the real axis, thus changing oscillatory terms into relaxation. We might hope that in the off-resonant case we are able to 
establish a similar crossover criterion, and again use this to investigate the effects of bath correlations and the cut-off frequency. However, we shall see that such an identification is less straightforward in the off-resonant regime. 

\begin{figure}[!t]
\begin{center}
\includegraphics[width=0.45\textwidth]{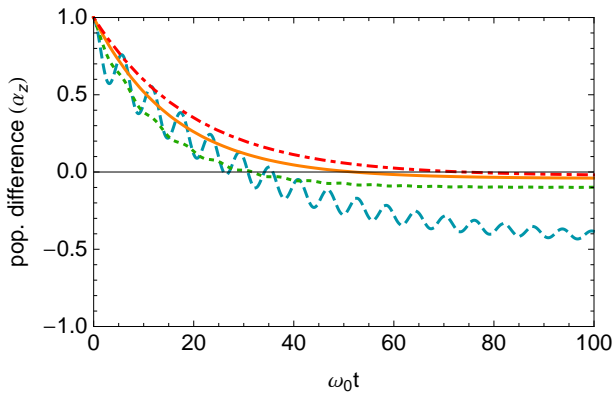}
\caption{Population difference for 
an off-resonant donor-acceptor pair 
as a function of scaled time $\w_0 t$. Temperatures $k_B T/\w_0=1$ (blue dashed curve), $k_B T/\w_0=5$ (green dotted curve),  $k_B T/\w_0=12$ (orange solid curve) and 
 $k_B T/\w_0=20$ (red dot-dashed curve) are shown. Parameters: $\alpha=0.05$, $V/\w_0=0.5$, $\epsilon/\w_0=1$, $\w_c/\w_0=4$, and $\mu=0.5$.}
\label{e-0d1/5_dynamics}
\end{center}
\end{figure}

We first present the full Bloch equations describing the evolution of our donor-acceptor pair for arbitrary energy mismatch. As in the resonant case, we have an equation of motion of the form
$\dot{\vec{\alpha}}=M\cdot\vec{\alpha}+\vec{b}$, but now the matrix $M$ is given by
\beq
M=\left(
\begin{array}{ccc}
-\Gamma_x & -(\epsilon+\lambda_1) & 0 \\
(\epsilon+\lambda_2) & -\Gamma_y & -2 B V_R \\
B^{-1} \zeta & B^{-1}(2 V_R+\lambda_3) & -\Gamma_z
\end{array}
\right),
\label{full_M}
\eeq
with $\vec{b}=( -B \kappa_x,-B\kappa_y,-\kappa_z)^T$. The rates become 
\begin{align}
\Gamma_x&=V^2(\gamma_{yy}(\eta)+\gamma_{yy}(-\eta))\label{Gamma_x},\\
\Gamma_y&=2V^2\left(\frac{4V_R^2}{\eta^2}\gamma_{xx}(0)+\frac{\epsilon^2}{2 \eta^2}(\gamma_{xx}(\eta)+\gamma_{xx}(-\eta))\right),
\end{align}
with $\Gamma_z=\Gamma_x+\Gamma_y$, and the energy shifts 
\begin{align}
\lambda_1&=\frac{2V^2\epsilon}{\eta}(S_{yy}(\eta)-S_{yy}(-\eta)),\\
\lambda_2&=\frac{2V^2\epsilon}{\eta}(S_{xx}(\eta)-S_{xx}(-\eta)),\\
\lambda_3&=\frac{4V^2V_R}{\eta}(S_{yy}(\eta)-S_{yy}(-\eta)).
\end{align}
The remaining quantities are 
\begin{align}
\zeta&=\frac{4V^2V_R \epsilon}{\eta^2}\left(\gamma_{xx}(0)-\frac{1}{2}(\gamma_{xx}(\eta)+\gamma_{xx}(-\eta))\right),\\
\kappa_x&=\frac{2V^2V_R}{\eta}(\gamma_{yy}(\eta)-\gamma_{yy}(-\eta)),\\
\kappa_y&=\frac{8V^2V_R \epsilon}{\eta^2}\left(S_{xx}(0)-\frac{1}{2}(S_{xx}(\eta)+S_{xx}(-\eta))\right),\\
\kappa_z&=\frac{V^2\epsilon}{\eta}\left((\gamma_{xx}(\eta)-\gamma_{xx}(-\eta))+(\gamma_{yy}(\eta)-\gamma_{yy}(-\eta))\right).
\label{kappa_z}
\end{align}
Here, $\eta=\sqrt{\epsilon^2+4 V_R^2}$ is the system Hamiltonian eigenstate splitting in the polaron frame. 

To exemplify the dynamics generated by the full off-resonant Bloch equations, 
in Fig.~\ref{e-0d1/5_dynamics} we plot the evolution of the population difference in the case of 
donor-acceptor energy mismatch, $\epsilon=2V$. 
By comparison of Fig.~\ref{e-zero_dynamics} (plotted in the resonant case) and Fig.~\ref{e-0d1/5_dynamics}, we see that 
the presence of a substantial 
energy mismatch causes 
the low-temperature population oscillations to increase in frequency but decrease markedly in amplitude, such that for $k_BT/\omega_0=5$ oscillations are now 
almost imperceptible. We also see that the population difference tends to a non-zero steady-state 
at low temperatures, as we might expect from simple thermodynamic arguments, since the states $\alpha_z=1$ and $\alpha_z=-1$ now have different energies. As the temperature is raised, however, the dynamics still looks to be approaching that shown in the resonant case of Fig.~\ref{e-zero_dynamics}. 


\subsection{Correlated fluctuations}

\begin{figure}[!t]
\begin{center}
\includegraphics[width=0.45\textwidth]{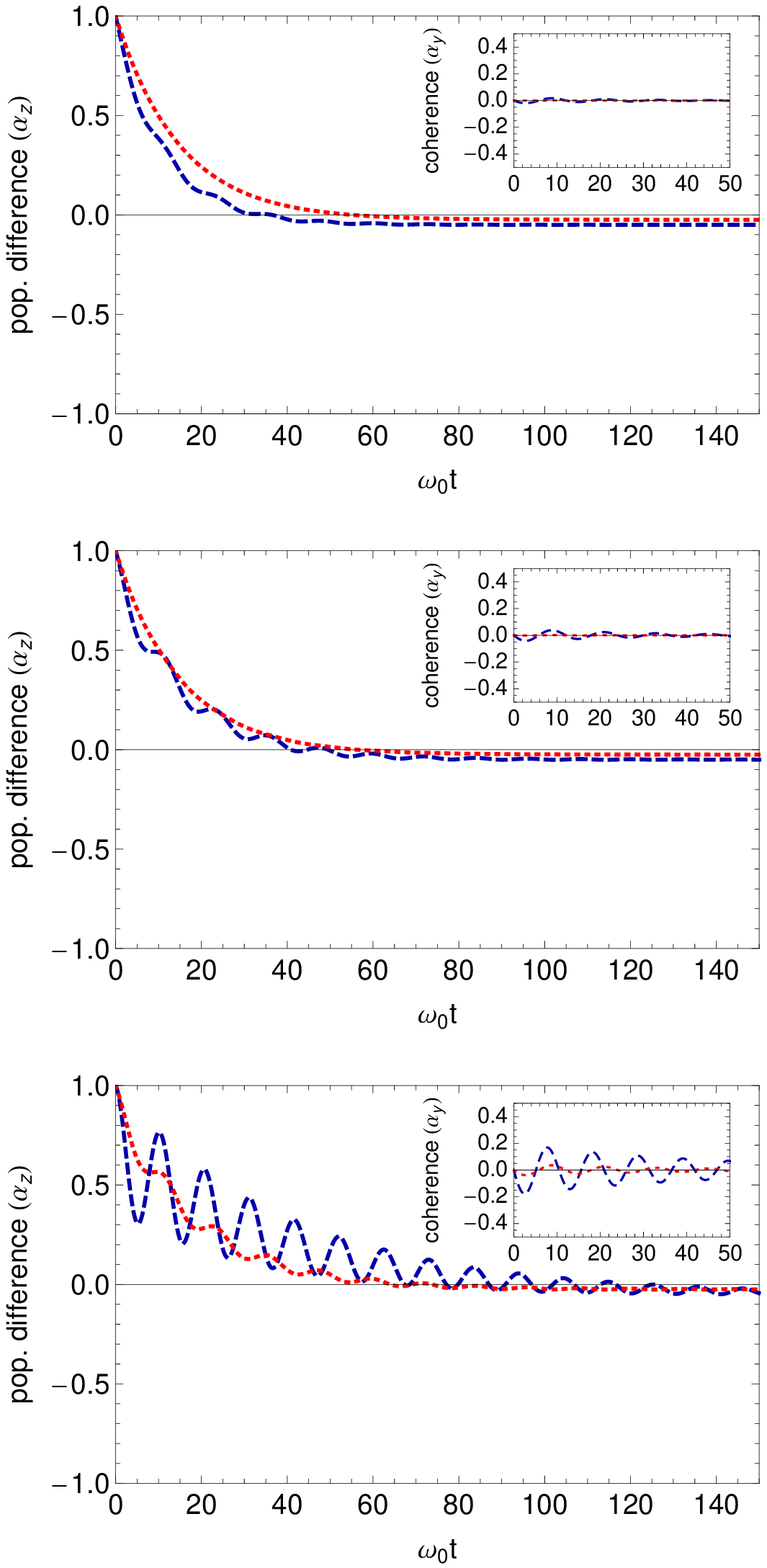}
\caption{Population difference as a function of scaled time $\w_0 t$ for temperatures 
of $k_B T/\w_0=5$ (blue dashed curve) and $k_B T/\w_0=10$ (red dotted curve), and for separations 
corresponding to no fluctuation correlations, $\mu=c/\w_0d=0$ (top), weak correlations, $\mu=0.5$ (middle) 
and strong correlations $\mu=2$ (botttom). The insets show the evolution of the 
corresponding coherence $\alpha_y$. Parameters: $\alpha=0.05$, $V/\w_0=0.5$, $\epsilon/\w_0=0.5$, and $\w_c/\w_0=4$.}
\label{off_resonant_dynamics}
\end{center}
\end{figure}

Let us now look at the effect of correlated fluctuations in the off-resonant dynamics. Though analysis of the 
full Bloch equations is now more complicated than in the resonant limit, we should still expect changes in the level of donor-acceptor fluctuation correlation 
to have a qualitatively similar effect on the transfer process as outlined in Section~\ref{resonant_transfer}. To illustrate that this is indeed the case, in Fig.~{\ref{off_resonant_dynamics}} we plot the donor-acceptor population dynamics under off-resonant conditions for three different levels of fluctuation correlation (increasing from top to bottom). Just as we found in the resonant case of Fig.~\ref{resonant_dynamics}, an increase in correlations enhances the lifetime of coherence present in the energy transfer process, and can even move the dynamics from a high temperature (or strong-coupling) predominantly incoherent regime to an {\it effective} low temperature (or weak-coupling) regime displaying pronounced coherent oscillations. In addition, in the off-resonant case stronger correlations also serve to amplify the coherent contribution to the full energy transfer dynamics (made up of distinct coherent and incoherent parts, as we shall show below), since the renormalised interaction strength $V_R$ increases in relation to the energy mismatch $\epsilon$. 

In an effort to put these qualitative observations on a more quantitive footing we could analyse the eigensystem of the full off-resonant $M$ [Eq.~(\ref{full_M})], in a similar manner to the resonant case.
However, finding the eigensystem of $M$ 
is now far less straightforward and analytical solutions to the full Bloch equations 
are consequently lengthy, and therefore of little direct use in gaining an understanding of the behaviour seen in Figs.~\ref{e-0d1/5_dynamics} and~{\ref{off_resonant_dynamics}}. 
The rest of this section is thus devoted to deriving simplified expressions for the energy transfer dynamics in two 
important limits: (i) weak-coupling, or strong correlations, where coherent dynamics can dominate, and (ii) high temperatures (and weak correlations), where the dynamics is similar in both the resonant and off-resonant cases. These expressions not only provide insight into the off-resonant behaviour of the system and the effect of correlated fluctuations, but also serve to highlight the difficultly in now defining a simple crossover criterion, as was possible in the resonant case.

\subsection{Weak-coupling (or strong correlation) limit}

We begin by considering the weak system-bath coupling limit, which we obtain by expanding all relevant quantities to first order in $J(\w)$. In fact, for strong enough fluctuation correlations this limit is attainable even if the system-bath coupling is not weak and/or the temperature not low, due to the factor $(1-F_D(\omega,d))$ appearing in Eqs.~(\ref{B_integral}) and~(\ref{phi}). 
With reference to our expressions for the correlation functions [Eqs.~({\ref{Lambda_xx}}) and 
({\ref{Lambda_yy}})], we see that within this approximation $\Lambda_{xx}(\tau)\rightarrow 0$ while 
$\Lambda_{yy}(\tau)$ remains finite. We may then set to zero all rates and energy shifts which are functions of $\Lambda_{xx}(\tau)$ only in Eq.~(\ref{full_M}). This results in the far simpler form
\beq
M_{W}=\left(
\begin{array}{ccc}
-\Gamma_{W} & -(\epsilon+\lambda_1) & 0 \\
\epsilon & 0 & -2BV_R \\
0 & B^{-1}(2V_R+\lambda_3) & -\Gamma_{W}
\end{array}
\right),
\eeq
where the weak-coupling rate is given by~\cite{rozbicki08}
\beq
\Gamma_{W}=4\pi \bigg(\frac{V_R}{\eta}\bigg)^2 J(\eta)(1-F_D(\eta,d))\coth(\beta \eta/2),
\label{off_resonant_weak_rate}
\eeq
and the two energy shifts may be written 
$\lambda_1=(\epsilon/\eta)\Lambda$ and $\lambda_3=(2 V_R/\eta)\Lambda$, with $\Lambda=2 V^2(S_{yy}(\eta)-S_{yy}(-\eta))$.
The inhomogeneous term becomes $\vec{b}_{W}=\{-B \kappa_x,0,-(\epsilon/2V_R)\kappa_x\}^T$ in the same limit, which leads to the weak-coupling steady state values of 
\begin{align}
\alpha_x(\infty)&=-\frac{2BV_R}{\eta} \tanh(\beta \eta/2),\label{sigma_x_steady_state}\\
\alpha_z(\infty)&=-\frac{\epsilon}{\eta}\tanh(\beta \eta/2),
\label{sigma_z_steady_state}
\end{align}
and $\alpha_y(\infty)=0$. 
As in the resonant case (in which there was no weak-coupling approximation), this steady-state has the same form as that expected from a standard weak-coupling approach, though with the replacement $V\rightarrow V_R$, and the extra factor of $B$ suppressing the coherence $\alpha_x(\infty)$. As the energy mismatch increases in relation to $V$, the weak-coupling steady state therefore becomes increasingly localised in the lower energy state $|2\rangle\equiv|GX\rangle$. Interestingly, this contrasts with the qualitatively incorrect form (at low temperatures at least) given by the Non-Interacting Blip Approximation (NIBA), $\alpha_z^{\rm NIBA}(\infty)=-\tanh{(\beta\epsilon/2)}$,~\cite{weissbook, leggett87} which predicts complete localisation in the lower energy state at zero temperature, regardless of the size of $\epsilon/V$. We should thus expect the present theory to fair far better than the NIBA for low-temperatures (or weak-coupling) in the off-resonant case, $\epsilon\neq0$. The rate $\Gamma_W$ given in Eq.~(\ref{off_resonant_weak_rate}) is also of the form expected from a weak-coupling treatment, though once more with the renormalisation  $V\rightarrow V_R$. In fact, such a replacement is sometimes made by hand in weak-coupling theories to provide agreement with numerics over a larger range of parameters,~\cite{rozbicki08} though it arises naturally in the polaron formalism here. We can therefore conclude that, in addition to allowing for the exploration of multiphonon effects,~\cite{jang08, nazir09, wurger98,rae02} the polaron master equation provides a rigorous way to explore the (single-phonon) weak-coupling regime for spectral densities of the type in Eq.~(\ref{spectral_density}).~\cite{wurger98}

As before, to find the time evolution of $\vec{\alpha}$ we evaluate the 
eigensystem of $M_W$ and use Eq.~({\ref{general_evolution}}). For the initial state $\vec{\alpha}(0)=\{0,0,1\}^T$ we find population dynamics 
%
\begin{eqnarray}
\alpha_z(t)&{}={}&\frac{\epsilon}{\eta}\bigg(\frac{\epsilon}{\eta}\e^{-\Gamma_{W} t}-\bigl(1-\e^{-\Gamma_{W} t}\bigr)\tanh(\beta\eta/2)\bigg)\nonumber\\
&&\:{+}\frac{4 V_R^2}{\eta^2}\e^{-\frac{\Gamma_{W} t}{2}}\bigg(\cos\Bigl(\frac{\xi_{\mathrm{W}} t}{2}\Bigr)-\frac{\Gamma_{W}}{\xi_{\mathrm{W}}}
\sin\Bigl(\frac{\xi_{\mathrm{W}} t}{2}\Bigr)\bigg),\nonumber\\
\label{weak_alphaz_t}
\end{eqnarray}
where the weak coupling oscillation frequency is given by 
\beq
\xi_{\mathrm{W}}=\sqrt{4\eta(\eta+\Lambda)-\Gamma_{W}^2},
\label{xi_OW}
\eeq
which we expect to be real to be consistent with our original expansion.
The first term in Eq.~({\ref{weak_alphaz_t}}), proportional to $(\epsilon/\eta)$ and present nowhere in the resonant case, describes incoherent 
relaxation towards the steady state value given by Eq.~({\ref{sigma_z_steady_state}}). The second term, proportional to 
$(V_R/\eta)^2$ and having a similar form to the resonant dynamics, describes damped oscillations with frequency $\xi_{W}$. Importantly, these oscillations have a temporal maximum amplitude of $4V_R^2/\eta^2\leq 1$, compared to $1$ in the resonant case. The effect of the energy mismatch in this limit is thus to suppress the amplitude of any oscillations in the population difference, while increasing their frequency due to the dependence of $\xi_W$ on $\eta$ in Eq.~(\ref{xi_OW}), exactly as seen in Fig.~\ref{e-0d1/5_dynamics}. 

\subsection{High temperature (or far from resonance) limit}

At high temperatures and weak correlations, we find that 
the population dynamics appears to be relatively insensitive to the size of the energy mismatch. In order to investigate this effect in more detail, we shall now make a high-temperature (or strong system-bath coupling) approximation to the full energy transfer dynamics.

Specifically, we consider the regime $V_R/\epsilon \ll 1$. This limit can in fact be achieved in two possible ways. Firstly, 
recalling that $V_R=BV$, we see that $V_R$ can be made small by increasing the system-bath coupling strength 
or temperature, such that $B \ll 1$. Alternatively, if the donor-acceptor pair are far from resonance, the ratio 
$V/\epsilon$ will be small, and hence $V_R/\epsilon$ smaller still. Observing that the correlation functions given 
by Eqs.~({\ref{Lambda_xx}}) and ({\ref{Lambda_yy}}) are both proportional to $B^2$, we can see that all 
dissipative terms in the equation of motion, $\dot{\vec{\alpha}}=M\cdot\vec{\alpha}+\vec{b}$, are at least of order $V_R^2$. We proceed by 
keeping only terms up to order $(V_R/\epsilon)^2$ in the full off-resonant $M$ and $\vec{b}$. This allows us to set $\lambda_3$, $\zeta$, $\kappa_x$ and $\kappa_y$ to 
zero, while the remaining quantities reduce to
\begin{align}
\Gamma_y&=V^2(\gamma_{xx}(\eta)+\gamma_{xx}(-\eta)),\\
\Gamma_z&=V^2\big(\gamma_{xx}(\eta)+\gamma_{xx}(-\eta)+\gamma_{yy}(\eta)+\gamma_{yy}(-\eta)\big),\\
\lambda_1&=2V^2(S_{yy}(\eta)-S_{yy}(-\eta)),\\
\lambda_2&=2V^2(S_{xx}(\eta)-S_{xx}(-\eta)),\\
\kappa_z&=V^2\big(\gamma_{xx}(\eta)-\gamma_{xx}(-\eta)+\gamma_{yy}(\eta)-\gamma_{yy}(-\eta)\big).
\end{align}
Hence, in the high-temperature limit, Eq.~(\ref{full_M}) takes on the simpler form
\beq
M_{\mathrm{HT}}=\left(
\begin{array}{ccc}
-(\Gamma_z-\Gamma_y)& -(\epsilon+\lambda_1) & 0 \\
(\epsilon+\lambda_2) & -\Gamma_y & -2BV_R \\
0 & 2B^{-1}V_R & -\Gamma_z
\end{array}
\right),
\eeq
while the inhomogeneous term reduces to $\vec{b}_{\mathrm{HT}}=\{0,0,-\kappa_z\}^T$. We then find the approximate steady-state population difference
\begin{align}
\alpha_z(\infty)=-\bigg(1+\frac{4 V_R^2}{\epsilon^2}\bigg(\frac{\Gamma_y}{\Gamma_z}-1\bigg)\bigg)\tanh(\beta \eta/2),
\end{align}
valid up to second order in $V_R/\epsilon$. For $V_R\ll\epsilon$, this steady-state is strongly localised in the low energy state ($\alpha_z(\infty)\approx-1$) if $\epsilon\gg k_BT$, though for $\epsilon\ll k_BT$ thermal effects dominate and $\alpha_z(\infty)\approx0$ as in the resonant case. Again, this behaviour tallies with Fig.~\ref{e-0d1/5_dynamics}.

To obtain the corresponding population dynamics, we note in reference to Eq.~(\ref{general_evolution}) that the coefficients $a_i$, the eigenvectors $\vec{m}_i$, and the eigenvalues $q_i$ will contain powers of our expansion parameter $V_R/\epsilon$. 
Expanding both $q_i$ and the products $a_i \vec{m}_i$ to second order, we find
\begin{eqnarray}
\alpha_z(t)&=&\e^{-\Gamma_z  t}\left(1-\frac{4V_R^2}{\epsilon^2}\right)+\frac{4V_R^2}{\epsilon^2}\e^{-\Gamma_z t/2}\cos(\bar{\epsilon} t)\nonumber\\
&&\:{-}(1-\e^{-\Gamma_z t})\tanh\left(\frac{\beta\eta}{2}\right)\left[1+\frac{4V_R^2}{\epsilon^2}\bigg(\frac{\Gamma_y}{\Gamma_z}-1\bigg)\right]\nonumber\\
\label{alpha_z_HT}
\end{eqnarray}
where the shifted oscillation frequency is
\beq
\bar{\epsilon}=\epsilon+(1/2)(\lambda_1+\lambda_2)+2 \epsilon(V_R/\epsilon)^2.
\label{barepsilon}
\eeq
As in the weak-coupling case [Eq.~(\ref{weak_alphaz_t})] the evolution of 
the donor-acceptor population difference consists of two contributions; incoherent relaxation towards the steady-state, and an oscillatory component with vanishing amplitude as $V_R/\epsilon\rightarrow 0$. The energy mismatch again serves to suppress oscillations in the population difference. 

The most striking feature, however, of Eq.~({\ref{alpha_z_HT}}) is that there is an oscillatory component at frequency $\bar{\epsilon}$ at all. In the high-temperature 
limit, we might expect that this frequency would reach a point where it becomes imaginary and $\alpha_z(t)$ displays 
purely incoherent relaxation, as in the equivalent resonant case. However, we can see that this is not the case since $\bar{\epsilon}$ is always real by definition. Furthermore, at very high temperatures $\bar{\epsilon}\rightarrow\epsilon$, and it therefore also remains finite. Eq.~({\ref{alpha_z_HT}}) thus highlights an important difference between the energy transfer dynamics in resonant and off-resonant situations. In the resonant case, as temperature is increased, the energy transfer process becomes 
less coherent through a reduction in oscillation frequency (i.e. $V_R$ becomes small in comparison to $\Gamma_z-\Gamma_y$), eventually reaching a point at which population relaxes incoherently towards the steady state. In the off-resonant case, the transfer process becomes less coherent 
predominately through a reduction in oscillation amplitude. For high temperatures, an oscillatory component is still (in theory) present in the system, although it becomes ever more dominated by incoherent relaxation towards the steady-state population distribution, which depends upon the ratio $\epsilon/k_BT$. These features are clearly seen in Fig.~\ref{e-0d1/5_dynamics}.

\begin{figure}[!t]
\begin{center}
\includegraphics[width=0.45\textwidth]{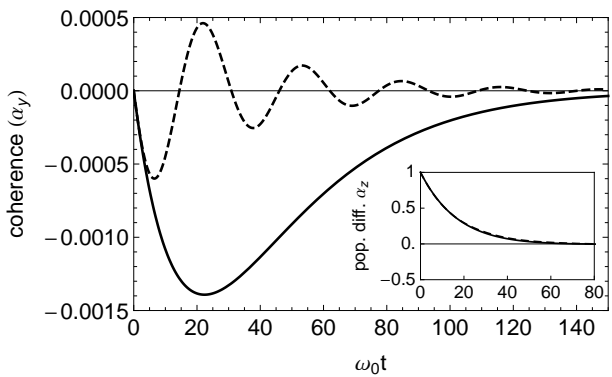}
\caption{Coherence ($\alpha_y$) as a function of scaled time $\w_0 t$ for resonant ($\epsilon=0$, solid curve) and off-resonant ($\epsilon/\w_0=0.2$, dashed curve) cases. The temperature, $k_B T/\w_0=13$, is chosen to be above the relevant crossover $T_c$ in the resonant case, such that the resonant dynamics is guaranteed to be incoherent. Parameters: $\alpha=0.05$, $V/\w_0=0.5$, 
$\w_c/\w_0=4$, and $\mu=0.5$. The inset shows the corresponding 
population dynamics.}
\label{smallcoherences}
\end{center}
\end{figure}

Only to first order in $V_R/\epsilon$ do our expressions predict purely incoherent off-resonant population transfer: 
\beq
\alpha_z(t)=\e^{-\Gamma_z  t}-(1-\e^{-\Gamma_z t})\tanh(\beta \eta/2).
\eeq
Let us also consider the evolution of 
$\alpha_y$ in the same limit: 
\begin{equation}
\alpha_y(t)=-\frac{2 B V_R}{\epsilon}\e^{-(1/2)\Gamma_z t}\sin(\bar{\epsilon} t).\label{alphayhighT}
\end{equation}
Hence, although the donor-acceptor population itself evolves entirely incoherently in this limit, the coherences may still perform oscillations due to the energy mismatch. To illustrate the difference in the transition to incoherent population transfer on- and off-resonance, in the main part of Fig.~\ref{smallcoherences} we plot the evolution of the coherence $\alpha_y(t)$ in both cases. The parameters have been chosen such that the resonant dynamics is in the incoherent regime ($T>T_c$), hence the resonant $\alpha_y$ displays no oscillations [see Eq.~(\ref{y_general_evolution})]. In accordance with Eq.~(\ref{alphayhighT}), however, the introduction of an energy mismatch induces oscillations in the donor-acceptor coherence. While these oscillations have an almost negligible amplitude, this behaviour serves to illustrate the subtlety in defining a strict crossover from coherent to incoherent dynamics in the off-resonant case. In particular, despite the different forms of coherence behaviour, the corresponding (essentially incoherent) population dynamics shown in the inset is almost indistinguishable in the two cases, even though there should still be a strongly suppressed coherent contribution in the off-resonant curve.

\begin{figure}[!t]
\begin{center}
\includegraphics[width=0.45\textwidth]{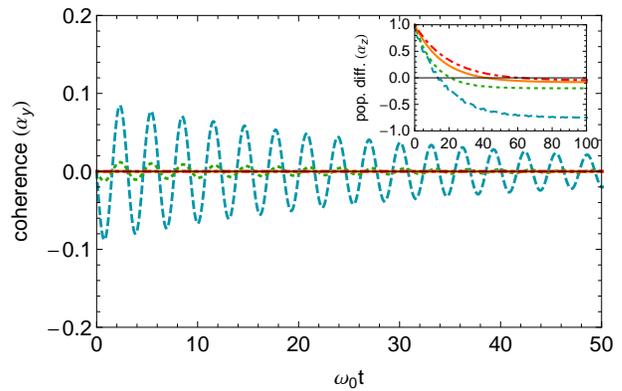}
\caption{Coherence ($\alpha_y$) as a function of scaled time $\w_0 t$ for temperatures 
of $k_B T/\w_0=1$ (blue dashed curve), $k_B T/\w_0=5$ (green dotted curve),  $k_B T/\w_0=12$ (orange solid curve) and $k_B T/\w_0=20$ (red dot-dashed curve). Parameters: $\alpha=0.05$, $V/\w_0=0.5$, 
$\w_c/\w_0=4$, $\epsilon/\w_0=2$ and $\mu=0.5$. The inset shows the corresponding 
population dynamics.}
\label{coherences}
\end{center}
\end{figure}

An alternative way to obtain oscillations of the coherence $\alpha_y$ in a regime of predominantly incoherent population transfer is to introduce a large energy mismatch (i.e. make $V/\epsilon$ small) at low temperature, as shown in Fig.~{\ref{coherences}}. Here, for the lowest temperature considered the population relaxes towards its steady state value with little sign of oscillation, while the coherence performs oscillations with a significant amplitude and considerable lifetime. This behaviour is strongly suppressed, however, as temperature increases, such that $k_BT>\epsilon$.

\section{Summary}
\label{summary}

Motivated by recent experiments which suggest that quantum coherence can survive in energy transfer processes even under potentially adverse environmental conditions,~\cite{lee07,engel07, calhoun09,collini09sc,collini09,collini10,panitchayangkoona10,mercer09,womick09} 
we have investigated various factors that determine the nature of the energy transfer dynamics in a model donor-acceptor pair. To do so, we used a polaron transform, Markovian master equation technique.~\cite{nazir09} This formalism is attractive as it allows for exploration of both the low-temperature (or weak-coupling) and high-temperature (or strong-coupling) regimes, as well as reliable interpolation between these two limits, provided the ratio $V/\omega_c$ does not become too large.~\cite{wurger98,jang09,rae02} We are also able to consistenly describe off-resonant effects, unlike in the NIBA,~\cite{leggett87,weissbook} and the influence of bath correlations. 

In the resonant case we identified a crossover temperature separating coherent and incoherent energy transfer. We found a non-trivial dependence of this temperature on both the degree of spatial correlation within the bath-induced fluctuations, and also on the cut-off frequency of the bath spectral density. Smaller cut-off frequencies were found to enhance the extent to which bath spatial correlations are able to protect coherence in the system. The crossover generally occurs in a high-temperature limit where multiphonon effects dominate, and so could not be captured by a standard perturbative treatment of the system-bath interaction. 

In the off-resonant case we found that coherent and incoherent regimes are less easily defined. In particular, 
for a sufficiently large energy mismatch between the donor and acceptor, coherence 
can in theory be present at all but infinite temperatures, albeit with an ever decreasing amplitude. However, using analytic expressions derived in various limits, we were able to characterise the off-resonant energy transfer process over much of the parameter space, illustrating the  suppression of coherence in the population dynamics with increasing temperature or energy mismatch. We also showed that strong correlations have a qualitatively similar 
effect to the resonant case, protecting coherence in the transfer process.

While we have concentrated in this work on elucidating general features of donor-acceptor energy transfer dynamics using a simple model system, the insight we have gained could be relevant to a variety of systems. In addition to those already mentioned,~\cite{lee07,engel07, calhoun09,collini09sc,collini09,collini10,panitchayangkoona10,mercer09,womick09} closely-spaced pairs of semiconductor quantum dots could provide a solid-state implementation of the model studied here.~\cite{gerardot05} In particular, our polaron master equation theory provides a bridge between the weak~\cite{rozbicki08} and strong~\cite{govorov05} system-bath coupling approximations already explored in this context. It would also be interesting to analyse the energy transfer dynamics of larger donor-acceptor complexes within the polaron formalism,~\cite{kolli10} to see if further understanding of the interplay between coherent and incoherent processes in such systems could be obtained. Finally, it would be desirable to perform a thorough investigation of the regime of validity of the polaron approach by comparison to numerically exact techniques.~\cite{makri98} 

\acknowledgements
We are very grateful to Alexandra Olaya-Castro, Andrew Fisher, and Avinash Kolli for interesting discussions and useful comments. This research was supported by the EPSRC and Imperial College London.

\appendix

\section{High temperature rates}
\label{high_temperature_rates}

Here we show how to obtain analytic approximations for the decoherence rates at high temperatures by 
use of a saddle point integration.  Within our formalism there are two rates which need to 
be evaluated:
\begin{align}
\gamma_{xx}( \eta)&=\frac{B^2}{2}\int_{-\infty}^{+\infty}\mathrm{d}\tau\e^{ i \tau \eta}(\e^{\phi(\tau)}+\e^{-\phi(\tau)}-2),\\
\gamma_{yy}( \eta)&=\frac{B^2}{2}\int_{-\infty}^{+\infty}\mathrm{d}\tau\e^{ i \tau \eta}(\e^{\phi(\tau)}-\e^{-\phi(\tau)}),
\end{align}
where $\phi(\tau)$ is given by Eq.~({\ref{phi}}). With the appropriate manipulations~\cite{wurger98} it is possible to show that
\begin{align}
\gamma_{xx}( \eta)&=\frac{B^2}{2}\e^{\beta \eta/2}\int_{-\infty}^{+\infty}\mathrm{d}\tau\e^{ i \tau \eta}(\e^{\tilde{\phi}(\tau)}+\e^{-\tilde{\phi}(\tau)}-2),\label{gamma_xx_tilde}\\
\gamma_{yy}(\eta)&=\frac{B^2}{2}\e^{\beta \eta/2}\int_{-\infty}^{+\infty}\mathrm{d}\tau\e^{ i \tau \eta}(\e^{\tilde{\phi}(\tau)}-\e^{-\tilde{\phi}(\tau)}),
\label{gamma_yy_tilde}
\end{align}
where now $\tilde{\phi}(\tau)=\tilde{\phi}(-\tau)=\phi(\tau-i\beta/2)$, and is given explicitly in integral form by
\beq
\tilde{\phi}(\tau)=2 \int_0^{\infty}\mathrm{d}\w\frac{J(\w)}{\w^2}(1-F_D(\w,d))\frac{\cos(\w \tau)}{\sinh(\beta \w/2)}.
\eeq
Using a super-Ohmic form of spectral density, $J(\w)=\alpha \w^3\w_0^{-2} \e^{-\w/\w_c}$, and 
assuming system-bath coupling in three 
dimensions such that $F_D(\w,d)=\mathrm{sinc}(\w d/c)$, allows $\tilde{\phi}(\tau)$ to be found analytically. We 
find $\tilde{\phi}(\tau)=\phi_0 C(x,y,\tau')$, where
\beq
\begin{split}
&C(x,y,\tau')=\\
&\frac{-i}{2 \pi x}\bigg[ \psi\left(\frac{1}{2}+\frac{1}{y}-\frac{i}{\pi}(\tau'+x)\right)-\psi\left(\frac{1}{2}+\frac{1}{y}-\frac{i}{\pi}(\tau'-x)\right)\\
&\,\,\,+\psi\left(\frac{1}{2}+\frac{1}{y}+\frac{i}{\pi}(\tau'-x)\right)-\psi\left(\frac{1}{2}+\frac{1}{y}+\frac{i}{\pi}(\tau'+x)\right)\bigg]\\
&+\frac{1}{\pi^2}\bigg[\psi'\left(\frac{1}{2}+\frac{1}{y}-\frac{i}{\pi}\tau'\right)-\psi'\left(\frac{1}{2}+\frac{1}{y}+\frac{i}{\pi}\tau'\right)\bigg].\nonumber\\
\label{tilde_phi}
\end{split}
\eeq
Here, $\phi_0=2 \pi^2 \alpha/(\w_0^2\beta^2)$, $x= \pi d/c \beta$, $y=\w_c \beta$, $\tau'=\pi \tau /\beta$, $\psi(z)$ is the the digamma function, and $\psi'(z)$ its first derivative.

To proceed, we assume a high-temperature or strong-coupling regime, such that the dominant contributions to the integrals in Eqs.~({\ref{gamma_xx_tilde}}) and ({\ref{gamma_yy_tilde}}) will come from the peak in $\tilde{\phi}(\tau)$ at $\tau=0$. More specifically, inspection of $C(x,y,\tau')$ reveals that for $y\gg1$ (the scaling limit of large $\omega_c$), we require $\phi_0\gg1$ for large $x$ (weak correlations), or $\phi_0x^2\gg1$ for small $x$ (strong correlations), in order for an expansion of $\tilde{\phi}(\tau)$ around $\tau=0$ to be valid. These definitions of the high-temperature (or strong-coupling) regime tally with the expansion parameters identified in Ref.~\onlinecite{nazir09}. In the opposite limit, $y\ll1$, we generally need $\phi_0x^2y^3/\pi^4\gg1$, except in the limit of very large separations (vanishing correlations), $x\rightarrow\infty$, where $\phi_0y\gg1$ is the relevant condition.

With these conditions in mind, we therefore expand $\tilde{\phi}(\tau)$ to second order in $\tau'=\pi \tau /\beta$, which gives 
\beq
\tilde{\phi}(\tau)\approx \phi_0(C_0(x,y)-\tau'^2C_2(x,y)),
\eeq
where
\beq
\begin{split}
C_0(x,y)=&\frac{i}{\pi x}\Big[\psi\Big(\frac{1}{2}+\frac{1}{y}+\frac{i x}{\pi}\Big)-\psi\Big(\frac{1}{2}+\frac{1}{y}-\frac{i x}{\pi}\Big)\Big]\\
&+\frac{2}{\pi^2}\psi'\Big(\frac{1}{2}+\frac{1}{y}\Big),
\label{C_0}
\end{split}
\eeq
and
\beq
\begin{split}
C_2(x,y)=&\frac{i}{2 \pi^3 x}\Big[\psi''\Big(\frac{1}{2}+\frac{1}{y}+\frac{i x}{\pi}\Big)-\psi''\Big(\frac{1}{2}+\frac{1}{y}-\frac{i x}{\pi}\Big)\Big]\\
&+\frac{1}{\pi^4}\psi'''\Big(\frac{1}{2}+\frac{1}{y}\Big).
\label{C_2}
\end{split}
\eeq
In this limit, Eqs.~(\ref{gamma_xx_tilde}) and~(\ref{gamma_yy_tilde}) are dominated by the terms containing a factor of 
$\exp[\tilde{\phi}(\tau)]$, allowing us to write
\beq
\gamma_{ll}(\eta)\approx \frac{B^2 \e^{\beta \eta /2}\beta}{2 \pi}\e^{\phi_0 C_0}\int_{-\infty}^{+\infty}d\tau'\e^{i \tau' \eta \beta/\pi}
\e^{-\tau'^2 \phi_0 C_2}.
\eeq
The integral is Gaussian and we arrive at the result
\beq
\gamma_{ll}(\eta)=\frac{\beta B^2\e^{\phi_0 C_0(x,y)}}{2 \sqrt{\pi C_2(x,y) \phi_0}}\e^{\beta \eta/2}\e^{-\beta^2\eta^2/(4 \pi^2 C_2(x,y)\phi_0)},
\eeq
which reduces to
\beq
\gamma_{ll}(\eta)\approx\gamma_{ll}(0)=\frac{\beta B^2\e^{\phi_0 C_0(x,y)}}{2 \sqrt{\pi C_2(x,y) \phi_0}},
\eeq
if the temperature is high enough such that $1/\eta \beta \gg 1$. 

It remains now to determine the bath renormalisation factor $B$, given by  Eq.~({\ref{B_integral}}). 
To do so it is helpful to separate vacuum and thermal contributions. 
We write $B=B_0 B_{\mathrm{th}}$, where
\beq
B_0=\exp\bigg[-\int_0^{\infty}d\w\frac{J(\w)}{\w^2}(1-F_D(\w,d))\bigg],
\eeq
and
\beq
B_{\rm th}=\exp\bigg[-\int_0^{\infty}d\w\frac{J(\w)}{\w^2}(1-F_D(\w,d))(\coth(\beta \w/2)-1)\bigg].
\eeq
Inserting the spectral density, and again assuming system-bath coupling in three dimensions, we find
\beq
B_0=\exp\bigg[-\alpha\frac{\w_c^2}{\w_0^2}\bigg(\frac{(d \w_c/c)^2}{1+(d \w_c/c)^2}\bigg)\bigg],
\label{B_vacuum}
\eeq
and
\begin{equation}
\begin{split}
B_{\rm th}=\exp\bigg[\frac{\phi_0}{2 \pi^2}&\bigg(\frac{i\pi}{x}\bigl(H(y^{-1}-i x/\pi)-H(y^{-1}+i x/\pi)\bigr)\\
&-2 \psi'(1+y^{-1})\bigg)\bigg],
\label{B_thermal}
\end{split}
\end{equation}
where $H(m)=\sum_{i=1}^m(1/i)$ is the $m$th harmonic number. We note that in the infinite separation (uncorrelated) limit, one finds $B_0(d\rightarrow \infty)=\exp[-\alpha(\w_c^2/\w_0^2)]$, and 
\beq
B_{\rm th}=\exp[\alpha(\w_c^2/\w_0^2)(1-y^{-2}(\zeta(2,1+y^{-1})+\zeta(2,y^{-1})))],
\eeq
where $\zeta(s,a)$ is the generalised Riemann zeta function.\\\\

\end{document}